\documentclass[12pt,journal,draftcls,letterpaper,onecolumn]{article}
\usepackage{wrapfig}
\usepackage{epsf}
\usepackage{graphicx}
\usepackage{subfigure}

\begin{document}

\title{Construction and Functional Analysis of Human Genetic Interaction Networks with Genome-wide Association Data}
\author{Gang Fang$^{1}\footnote{Corresponding author; Supplementary material: http://vk.cs.umn.edu/humanGI}$, Wen Wang$^{1}$, Vanja Paunic$^{1}$, \\ 
Benjamin Oately$^{1}$, Majda Haznadar$^{2}$, Michael Steinbach$^{1}$, \\
Brian Van Ness$^{2}$, Chad L. Myers$^{1}$, Vipin Kumar$^{1}$\\
$^{1}$Department of Computer Science, \\
$^{2}$Department of Genetics, Cell Biology, and Development\\
University of Minnesota, Minneapolis, MN 55455, USA.}

\date{}


\maketitle

\begin{abstract}

Genetic interaction measures how different genes collectively contribute to a phenotype, and can reveal functional compensation and buffering between pathways under genetic perturbations. Recently, genome-wide investigation for genetic interactions has revealed genetic interaction networks that provide novel insights both when analyzed independently and when integrated with other functional genomic datasets. For higher eukaryotes such as human, the above reverse-genetics approaches are not straightforward since the phenotypes of interest for higher eukaryotes such as disease onset or survival, are difficult to study in a cell based assay.

In this paper, we propose a general framework for constructing and analyzing human genetic interaction networks from genome-wide single nucleotide polymorphism (SNP) datasets used for case-control studies on complex diseases. Specifically, we propose a general approach with three major steps: (1) estimating SNP-SNP genetic interactions, (2) identifying linkage disequilibrium (LD) blocks and mapping SNP-SNP interactions to LD block-block interactions, and (3) functional mapping for LD blocks. We performed two sets of functional analyses for each of the six case-control SNP datasets used in the paper, and demonstrated that (i) genes in LD blocks showing similar interaction profiles tend to be functionally related, and (ii) the network can be used to discover pairs of compensatory gene modules (between-pathway models) in their joint association with a disease phenotype. The proposed framework should provide novel insights beyond existing approaches that either ignore interactions between SNPs or model different SNP-SNP pairs with genetic interactions separately. Furthermore, our study provides evidence that some of the core properties of genetic interaction networks based on reverse genetics in model organisms like yeast are also present in genetic interactions revealed by natural variation in human populations.

\end{abstract}

\section{Introduction}
\label{sec:intro}

Genetic interaction measures how different genes collectively contribute to a phenotype, and can reveal functional compensation and buffering between pathways (or protein complexes) under genetic perturbations. Recently, genome-wide screening of genetic interactions have become possible for different species via high-throughput methods \cite{costanzo2010genetic} in which the phenotypic effect of the double-knockout of each pair of genes are compared with the aggregated effects of the two individual knockouts under an assumption of independence. An extreme case, called synthetic lethality, occurs when a double knockout results in the death of a cell even when the single knockouts have no effect. The genetic interaction networks revealed by these experiments provide novel insights both when analyzed by themselves \cite{tong2004global} and when integrated with other molecular networks or genomic datasets \cite{costanzo2010genetic}, such as physical interaction \cite{kelley2005systematic}, gene expression \cite{hescott2010evaluating} and chemical-genetic interaction data \cite{parsons2003integration}. For higher eukaryotes such as human, these reverse-genetics approaches have not been as straightforward due to both less amenable genetics and more complex phenotypes of interest such as disease onset and survival, which are difficult to study in cell based assays. 

Despite slow progress in mapping genetic interactions using reverse genetics approaches in human cells, we are accumulating a wealth of data on individual genetic variations in human populations.  Genome-wide association studies (GWAS) capturing single-nucleotide polymorphisms (SNPs) or copy number variations \cite{moore2010bioinformatics} that have been widely applied for studying genetic differences between disease samples (cases), and normal samples (controls) \cite{hirschhorn2005genome}, offer an alternative means to map genetic interactions. For example, if two genetic variants have weak individual association with a disease but very strong joint association, the genes controlled by the two variants may have compensating functions that can buffer variations in each other, but yield a much higher risk of the disease phenotype of interest for a joint mutation.

Indeed, the genetic interactions between genetic variants (such as SNPs and CNVs) have been extensively studied as statistical epistasis (used interchangeable with genetic interaction in this paper) \cite{cordell2002epistasis} in both genome-wide \cite{cordell2009detecting} and targeted studies \cite{brian2008bmc,fang2010miningtkde,pcd2010tr}. However, existing approaches mostly consider pairs (or high-order combinations) of genetic variants as separate interaction candidates, and hence estimate their statistical significance and study their biological interpretations in an isolated manner \cite{cordell2009detecting}. While many statistically significant and biologically relevant instances of epistasis are discovered, these approaches may have the following drawback. Several pairs of genetic variants may not have statistically significant genetic interactions when considered as individual interaction candidates, but nevertheless, can be collectively significant if they are highly enriched for two pathways that have complementary functions (a between pathway model, BPM \cite{kelley2005systematic}). This limitation motivates the design of approaches that can model the collective significance of SNP pairs based on their network structure.

Specifically, we aim to construct human disease-specific (cases vs. controls) genetic interactions from GWAS case-control datasets and then discover BPMs from the constructed network, i.e., two sets of genetic variants that have many genetic interactions across the two sets but none or very few within either set. For the community doing research on genetic interaction, the novelty is the exploration of whether the biologically interesting structures (e.g., BPMs) discovered from the genetic interaction networks of lower eukaryotes such as yeast also exist in complex diseases, such as cancer and neurological diseases, for higher eukaryotes such as humans. For the community doing research on GWAS data analysis, this work provides additional evidence to support the shift from the analysis of single genes (or SNPs) to sets of genes (e.g., gene sets \cite{subramanian2005gsa,wang2007pathway} or protein interaction subnetworks \cite{subnetwork2007,baranzini2009pathway,moore2010bioinformatics}). Existing approaches that focus on the discovery of individually significant pathways or subnetworks ignore those pathways or subnetworks that are individually insignificant but are compensatory to each other as a combination (e.g., pairs of pathways or subnetworks) in their strong association with a disease phenotype.

In this paper, we propose a general framework for constructing human disease-specific genetic interaction networks with GWAS data. Because different types of genomic data have unique characteristics that need to be addressed, we focus on genome-wide case-control SNP data and its accompanying linkage disequilibrium (LD) structure. We discuss the challenges in the construction of genetic interaction networks due to LD structure and propose a general approach with three steps: (1) estimating SNP-SNP genetic interactions, (2) identifying LD blocks and summarizing SNP-SNP interactions to LD block-block genetic interactions, and (3) functional mapping (e.g. gene mapping) for each LD block. 

To illustrate how the constructed genetic interaction network can be used to obtain both known and novel biological insights about disease phenotype of interest in the case-control study we designed two sets of functional analyses ($I$ and $II$) to analyze the genetic interaction networks constructed on each of the six case-control SNP datasets used in this study.

For functional analysis $I$,  we study whether a constructed human genetic interaction network has functional significance with respect to independent biological databases. Specifically, we compare the LD block-block genetic interaction network and the genetic-interaction-profile-based LD block-block similarity network with the human functional network integrated in \cite{huttenhower2009exploring}. Interestingly, we find that the pairs of LD blocks that have high genetic interaction and those pairs that have high similarity of genetic interaction profiles have significantly higher functional similarity. This motivates the potential utility of the constructed genetic interaction network for revealing both known and novel biological insights into the disease phenotype of interests in a case-control study.

For functional analysis $II$, we study how to use the constructed human genetic interaction network to provide detailed insights about the compensation between pathways in their joint association with a disease phenotype. Specifically, we discover between pathway models (BPM) from the block-block genetic interaction network. A BPM contains two sets of LD blocks, which have many cross-set genetic interactions but very few within-set genetic interactions. The experiments on the six SNP datasets demonstrate that the discovered BPMs have statistically significant properties (supported by permutation tests of case-control groupings) such as across-set densities of genetic interactions and functional enrichments based on three sets of biological databases. The significant BPMs may provide indications of the compensation between pathways (or protein complexes) in their association with the disease phenotypes, and serve as a novel type of biomarker for complex diseases.

Comprehensive experimental results on the six case-control SNP datasets support several points:  (i) From the perspective of genetic interaction analysis, the constructed human genetic interaction network has functional significance, and the biologically interesting motifs such as BPM that are common in lower eukaryotes also exist in human with respect to complex diseases such as cancer and Parkinson's disease; (ii) From the perspective of GWAS data analysis and biomarker discovery, discovering BPMs from the constructed human genetic interaction network can help reveal novel biological insights about complex diseases, beyond existing approaches for GWAS data analysis that either ignore interactions between SNPs, or model different SNP-SNP genetic interactions separately rather than studying global genetic interaction networks as in this study.

\section{Methods}
\label{sec:methods}

We first present a general framework for constructing genetic interaction networks from genome-wide case-control SNP datasets, and then describe two approaches for the functional analysis of the resulting networks that can be used to discover novel biological insights about complex diseases. For each step in this process, we selected a particular approach. Various alternatives are possible, but due to the limitation of space and the desire to clearly explain our general approach, we do not discuss those alternatives in any detail. However, given the significant results obtained by the current approach consistent over six datasets (Section \ref{sec:exp}), some of these alternatives should be explored to see if additional improvements in the results are possible. 
	
\subsection{Network Construction}
\label{sec:construction}

There are three steps in the network construction framework: (i) measuring all pairwise SNP-SNP genetic interaction with respect to the case-control grouping in a GWAS dataset, (ii) summarizing SNP-to-SNP interactions into a block-level genetic interaction network, and (iii) functional mapping for each block.

\subsubsection{SNP-to-SNP Genetic Interaction}
\label{sec:synergy}

The principal goal of measuring genetic interactions between two SNPs is to capture the non-additive effect between the two SNPs in their combined association with the phenotype of interest. For this purpose, we leverage the extensive research on statistical epistasis \cite{cordell2002epistasis} that has been recently reviewed by H. Cordell \cite{cordell2009detecting}.

Among the different measures for epistasis, we selected the information theoretic \emph{synergy} \cite{anastassiou2007computational}. The \emph{synergy} between two SNPs with respect to a binary class label variable (cases vs. controls), i.e., $Syn_C(S_i,S_j)$ is defined in \cite{anastassiou2007computational} as follows:

\begin{equation}
Syn_C(S_i,S_j) = I(S_i,S_j;C) - I(S_i;C) - I(S_j;C),
\label{eq:synergy}
\end{equation}

where $I(X;C)$ denotes the mutual information between the class variable $C$ and a variable $X$

\begin{equation}
I(X;C) = \sum_{c \in C}\sum_{x \in X}p(x,c)log\left(\frac{p(x,c)}{p(x)p(c)}\right)
\label{eq:mi}
\end{equation}

In this paper, \emph{synergy} and mutual information are always normalized by $H(C)$, after which, $Syn_C(S_i,S_j)$ ranges from $-1$ to $+1$. The focus of this paper is on positive \emph{synergy} since we want to measure the interaction between two SNPs beyond the independent additive effect of their joint association with a case-control phenotype. The interpretation of a positive \emph{synergy}, e.g. 0.2, is that the two SNPs as a combination provide 20\% more information about $C$ beyond the summation of information provided independently by the two SNPs. In this paper, we say two SNPs have a positive genetic interaction if they have a positive \emph{synergy} (beyond-additive effect) to keep the sign of genetic interaction \emph{synergy} consistent. Note that, in reverse-genetics based yeast genetic interaction, negative genetic interaction is used to denote beyond-additive effect.

\begin{figure}[t]
\centering
\includegraphics[width=.95\textwidth]{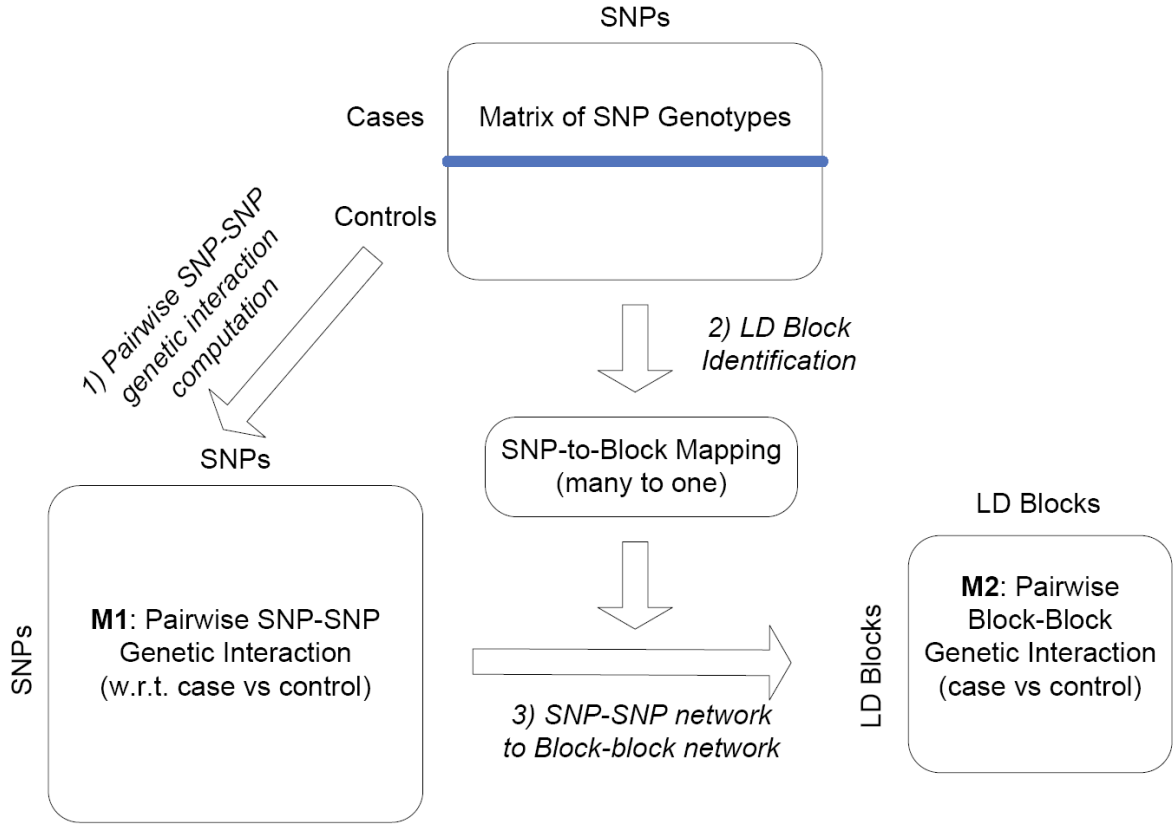}
\caption{Flowchart of constructing genetic interaction networks from case-control datasets. Three steps: (1) Calculating pairwise SNP-SNP genetic interactions, yielding $M1$; (2) Identifying LD blocks; and (3) summarizing a SNP-SNP Network to a block-block genetic interaction network ($M2$).} 
\label{fig:flowchar_construct}
\end{figure}

We chose to compute synergy for all pairs of SNPs rather than just those pairs for which one SNP has sufficiently large marginal effects \cite{hannum2009genome} since we did not want to risk missing SNP pairs that have weak (or no) marginal effect but a strong combined effect as discussed in \cite{zhang2010bayesian} since these pairs are essential for building an interaction network that may have even better statistical power than other approaches. After the \emph{synergy} calculation for all pairwise SNPs, we get a full weighted SNP-SNP network. We denote this matrix as $M1$, as shown in Figure \ref{fig:flowchar_construct}). This network cannot be directly interpreted because of the LD structure in the SNP data. In the next section, we discuss this challenge in detail and present an approach to address it.

\subsubsection{LD Block-block Genetic Interaction Network}
\label{sec:blockginet}

\begin{figure}[!t]
\centerline{\includegraphics[width=.95\linewidth]{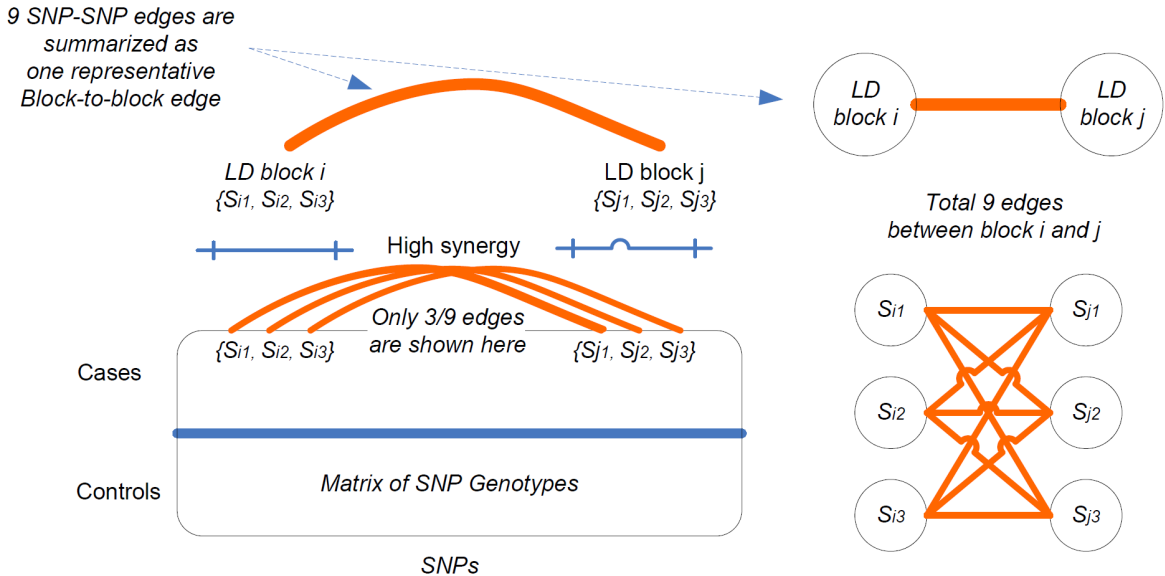}}
\caption{Illustration of a trivial biclique pattern due to LD structure in SNP data and how it is summarized into a representative block-block edge. Such biclique does not reflect the functional similarity between the SNPs in the same LD block. There should be only one edge connecting two functional units.} 
\label{fig:ld_biclique}
\end{figure}

Due to the LD structure in SNP data, nearby SNPs tend to have correlated genotypes over the samples. Therefore, if a pair of SNPs ($S_i$ and $S_j$) have large \emph{synergy}, the SNPs close to $S_i$ probably also have large \emph{synergy} with the SNPs close to $S_j$. This can result in a trivial type of local motif (approximate bicliques) in the SNP-SNP network as illustrated in Figure \ref{fig:ld_biclique}.This biases the functional analysis since such bicliques do not reflect the functional similarity between the SNPs in the same LD block. In order to gain non-trivial insights from the genetic interaction network, we propose to summarize the SNP-SNP interaction network by an LD block-block network.

\textbf{Identifying LD blocks}: Different measures for estimating LD such as $r^2$ and $D'$ \cite{dawson2002first} are specifically designed for measuring the non-random associations between polymorphisms at different loci. These measures capture the difference between observed and expected allelic frequencies (assuming random distributions), which depend on the phase information and define an LD block within a genomic region. With a related but different purpose, our goal in this study is to identify a set of SNPs on the same chromosome having similar genotype profiles and use a single block to represent them. 

We use Hamming similarity to measure the correlation between the genotype profiles of SNPs. (SNPs are columns in the case-control data as illustrated in Figure \ref{fig:flowchar_construct}.) This similarity serves better for our purpose of measuring mathematical similarity between two SNPs rather than the LD. We take such a conservative approach in order to make sure that we do not create two separate LD blocks for SNPs that have similar genotype profiles. This also avoids the estimation of phase information, which adds additional uncertainty that may confound the analysis.  We do not restrict an LD block to be within a local genomic region. This is because SNPs that are far from each other can also have high genotype correlation, at least from the mathematical perspective as shown in \cite{lawrence2005prospects}. Again, we take such a conservative approach in order to make sure that we do not create two separate LD blocks for SNPs that have similar genotype profiles, which may happen if a window-size constraint is used.  For simplicity, we perform a greedy search of LD blocks. Specifically, we randomly take a SNP and combine it with all the other SNPs (on the same chromosome) with Hamming similarity above a threshold $h$ ($h = 0.7$ is used in this study) as an LD block. A SNP will only be assigned to one LD block.

\textbf{SNP-SNP Network to LD Block-Block Network}:

After identifying all the LD blocks in a dataset, we have a many-to-one mapping from all the SNPs to a set of blocks. Given this mapping, we summarize the SNP-SNP \emph{synergy} network to a block-block \emph{synergy} network using the following general function to estimate the \emph{synergy} between two blocks $B_i$ and $B_j$,

\begin{equation}
Syn_C(B_i,B_j) = \Psi_{S_i \in B_i, S_j \in B_j}Syn_C(S_i,S_j)
\label{eq:synergyBlock}
\end{equation}

where $\Psi$ denotes a general aggregation function, e.g. $max$ or $mean$. In this paper, we adopt the $max$ function based on the following reasons and observations: (1) Biologically, it is likely that only one pair of SNPs across two LD blocks are truly causative in the case-control phenotype, in which case $max$ is the ideal aggregation function. (2) Based on multiple datasets used in the experiment section, the $max$ function consistently yields coherence with existing biological knowledge gained from yeast genetic interaction networks. (3) In the sanity check, the pairs with top \emph{synergy} values have similar LD-block sizes as the null distribution, and thus are not due to the bias of large LD block sizes. There are other aggregation approaches \cite{hannum2009genome, stram2004tagismb} that we will explore in future work.

After this step, we have a block-block genetic interaction network, which we denote as $M2$ as illustrated in Figure \ref{fig:flowchar_construct}.

\textbf{Functional Mapping for each LD Block}: After the construction of an LD block-block genetic interaction network, a functional mapping for each block is required to interpret the structure of the interaction network in functional terms that have biological meaning. For this purpose, we first assign each SNP to the closest gene based on its genome location. Then, the genes of an LD block are obtained from the SNPs that were assigned to that block in the LD identification step. We will explore more advanced gene mapping strategies in future work.
 Interestingly, even with this simple gene mapping approach, the functional analysis in Section \ref{sec:exp} shows that the constructed LD block-block interaction network still appears to have functional structure. Note that the gene mapping does not result in a gene-gene interaction network. This is because an LD block can span multiple genes so that gene-gene network derived from the blocks would contain many trivial biclique patterns. From this perspective, the genetic interaction network constructed for yeast in \cite{hannum2009genome} from eQTL data may have included a large number of false positive gene-gene edges since it connected all the gene pairs from the two LD blocks. In contrast, the block-block network constructed in this study has the least amount of bias from trivial bicliques. 

\subsection{Network Analysis}
\label{sec:analysis}

\begin{figure}[t]
\centering
\includegraphics[width=.98\textwidth]{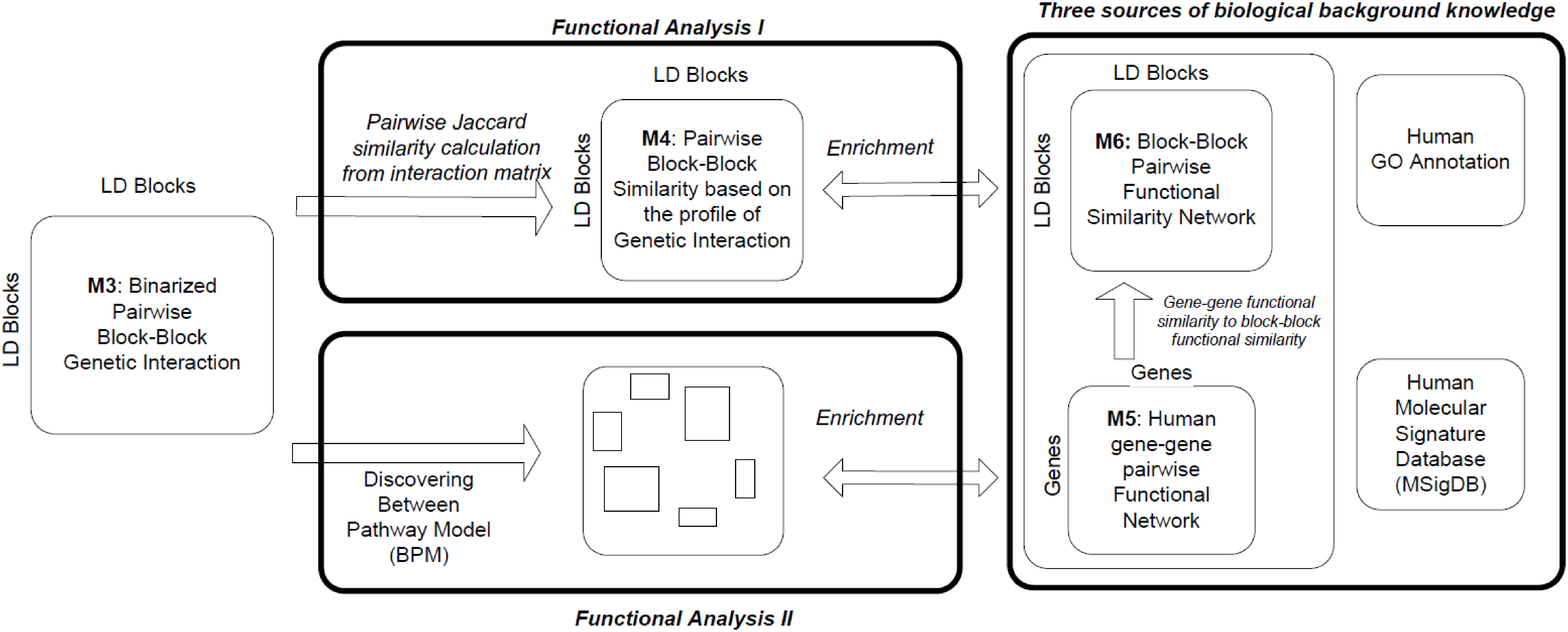}
\caption{Flowchart of the functional analysis on the binarized genetic interaction profiles $M3$. There are two types of analysis: ($I$) constructing the block-block similarity network ($M4$) from $M3$ and enrichment analysis, and ($II$) discovering BPMs (between pathway models) from $M3$ and enrichment analysis. Both enrichment analyses are with respect to three sources of biological knowledge, namely (i) the block-block functional similarity network ($M6$) derived from  the human functional network $M5$, (ii) the human GO annotations and  (iii) the human molecular signature database (MSigDB).} 
\label{fig:flowchart_bpm_net_analysis}
\end{figure}

In Section \ref{sec:construction}, we presented a framework for constructing a block-block genetic interaction network from a human case-control genomic dataset. In this subsection, we present two sets of functional analyses: $I$, which Comparing the constructed LD block-block genetic interaction network and the corresponding similarity network derived from the human functional network, and $II$, which discovers between pathway models (BPM) from the LD block-block genetic interaction network for functional enrichment analysis \cite{kelley2005systematic}. Both types of analysis have been used in the systematic interpretation of double-knockout-based yeast genetic interaction networks \cite{costanzo2010genetic,kelley2005systematic}. Here we use a similar approach to reveal novel biological insights from the genetic interaction networks constructed from human case-control genomic datasets.  Figure \ref{fig:flowchart_bpm_net_analysis} shows the overall design of the two types of functional analysis and will be referenced extensively in the rest of this section. 

An important point in understanding following discussion is that we first threshold the block-block genetic interaction matrix ($M_2$) to a binary matrix ($M_3$). Specifically, we binarize this network with a quantile threshold $q$ (e.g. 1\%), such that those block-block edges with \emph{synergy} in the top $q$ quantile (those with large beyond-addtive effect interactions) are kept in the binary network. We denote the matrix representation of this network as $M3$ as illustrated in Figure \ref{fig:flowchart_bpm_net_analysis}. Both of the analyses make use of binarized matrix $M3$.

\subsubsection{Functional analysis $I$: Comparing the constructed LD block-block genetic interaction network and the corresponding similarity network derived from a human functional network}

In the first approach, we study whether the constructed human genetic interaction network has functional significance supported by independent biological databases. Specifically, we first use Jaccard similarity to measure the similarity between the profiles of genetic interactions of two blocks in $M_3$. This results in matrix $M_4$ as shown in Figure \ref{fig:flowchart_bpm_net_analysis}. The motivation is that the analysis of global yeast genetic interaction networks \cite{costanzo2010genetic} has shown that the similarity between such interaction profiles of two genes is correlated with the functional similarity of two genes. In the experimental section, we will use the constructed block-block similarity matrix $M_4$ to compare the similarity of block interaction profiles with the human gene-gene functional network integrated in \cite{huttenhower2009exploring}.

\subsubsection{Functional analysis $II$: Discovering BPMs (between pathway models) from the LD block-block genetic interaction network for functional enrichment analysis}
\label{sec:bpm}

Functional analysis approach $II$, takes a complementary approach to discover between pathway models (BPM). (This approach was demonstrated to be effective in the analysis of yeast genetic interaction networks \cite{kelley2005systematic}.) Using insights gained from yeast genetic interaction network, a BPM contains two sets of genes that have many across-set genetic interactions and within-set protein-protein interactions \cite{kelley2005systematic}. The two sets of genes in a BPM may correspond to two biological pathways (or protein complexes) that have redundant (or complementarity) biological functions with respect to the case-control grouping. In the context of this study, a BPM contains two sets of LD blocks, which have many large cross-set genetic interactions (synergy values) in $M_3$ and very few within-set genetic interactions.

Discovering BPMs from the binarized disease-specific genetic interaction matrix ($M_3$) can provide novel insights beyond existing approaches designed for analyzing case-control SNP datasets. Recently, existing approaches for analyzing case-control datasets have shifted from discovering single genes (or SNPs etc) to sets of genes (e.g. pathways \cite{subramanian2005gsa} or protein interaction subnetworks \cite{subnetwork2007}). While many statistically significant and biologically relevant gene sets or subnetworks are discovered, existing approaches may ignore those pathways or subnetworks that are individually insignificant but are compensatory to each other as a combination in their strong association with a disease. From this perspective, a BPM captures the compensation between two pathways in their combined association with complex diseases such as cancer that may be caused by the perturbation of multiple (e.g., a pair of) pathways but not individual pathways. This may lead to the discovery of a new type of complex biomarker, i.e. pairs of compensating pathways or protein complexes.

Many algorithms have been proposed to discover BPMs from yeast genetic interaction networks. These algorithms mostly depend on integration of both physical interaction network and genetic interaction network \cite{kelley2005systematic,ulitsky2008maps}. Such integrative approaches have the advantage of better interpretability because of the integration with physical interaction data. However, given our goal of discovering BPMs from human case-control datasets and the fact that human protein interaction network is not as complete and lacks reproducibility \cite{huttenhower2009exploring}, an integrative approach may miss many BPMs that are not yet well-supported by the existing limited functional knowledge of the human genome.

For the above reason, in this study, we took an exploratory approach in which, we search for BPMs only from the genetic interaction matrix $M_3$. Given a binary symmetric matrix ($M_3$), the BPM discovery problem can be reformulated as a quasi-biclique discovery problem \cite{liu2010modeling}.

A quasi-biclique is defined as a non-weighted bipartite graph $G = (X \cup Y, E)$ (where $X$ and $Y$ are two sets of LD blocks and $E$ is a set of LD block-block genetic interaction edges) such that, for a given parameter $0 < \delta < 0.5$.

\begin{equation}
\forall x \in X, d(x,Y) \geq (1-\delta)|Y|,\ \\
\forall y \in Y, d(y,X) \geq (1-\delta)|X|,
\label{eq:quasibclique}
\end{equation}

where $d(x,Y)$ denotes the number of edges between $x$ and all the nodes in $Y$(similarly for $d(y,X)$). In this paper, we adopt a greedy algorithm with polynomial time complexity \cite{liu2010modeling} to efficiently search for quasi-blciques from the binary block-block network ($M_3$). Note that, several other algorithms that are designed for quasi-biclique discovery or BPM discovery \cite{ma2008mapping} can also be applied for the same purpose. These will be explored in future work as this paper focuses on presenting the overall framework.

Based on the definition, a quasi-biclique may also have many edges within each of the two sets ($X$ and $Y$), while BPMs that with no or very few within-set edges are relatively more interesting. Therefore, after the discovery of a set of quasi-bicliques, a postprocessing will be applied to further select a subset of bicliques with a small fraction of within-set genetic interactions as BPM candidates for the following functional analysis. We will design experiments to interpret the discovered BPMs with human functional network \cite{huttenhower2009exploring} and conduct enrichment analysis with human GO annotations \cite{ashburner2000gene} and human molecular signature database (MSigDB\cite{subramanian2005gsa}) in Section \ref{sec:bpmresults}.

\section{Results}
\label{sec:exp}

In this section, we present the experimental results on the two sets of functional analyses (described in Section \ref{sec:analysis}) on the network constructed with the framework presented in Section \ref{sec:construction}.

\subsection{Data Sets}

In the experiments, we use six case-control SNP datasets, one from a genome-wide association study on Parkinson's disease (disease vs. normal) \cite{fung2006genome}, and the others from targeted studies (with around 3000 SNPs over about 1000 genes) on Myeloma long vs. short survival \cite{brian2008bmc}, Myeloma cases vs. controls \cite{pcd2010bio}, lung cancer cases vs. controls (both heavy smokers) \cite{church2009prospectively}, rejection vs. no-rejection after kidney transplant \cite{he2009power}, bone disease (large vs. small number of bone lesions ) \cite{brian2008bmc}. We denote these 6 datasets as Parkinson, Myeloma-survival, Myeloma, Lung, Kidney, Bone in this section. For the genome-wide Parkinson data, we selected 8994 non-synonymous SNPs for this study for the concern of computational efficiency, given the huge number of all pairwise SNP-SNP genetic interaction to compute, especially in a large number (100) of permutations for each dataset (functional analysis II).

Table \ref{tab:datasets} displays the information about the six datasets, including the number of cases, controls, SNPs, LD blocks identified and edges in an LD block-block genetic interaction network ($M_3$).

\begin{table}[!t]
\centering
\begin{tabular}{llllll}\hline
Dataset & \ensuremath{\sharp} of & \ensuremath{\sharp} of & \ensuremath{\sharp} of & \ensuremath{\sharp} of LD & \ensuremath{\sharp} of B-B \\
 & Cases & Controls & SNPs & Blocks & Edges \\\hline
Parkinson & 270 & 271 & 8994 & 5011 & 31387 \\
Myeloma-survival & 70 & 73 & 3077 & 986 & 7291 \\
Lung & 99 & 99 & 2990 & 977 & 7159 \\
Kidney & 135 & 136 & 2864 & 963 & 6955 \\
Myeloma & 244 & 149 & 2345 & 804 & 4848 \\
Bone & 73 & 185 & 2982 & 993 & 7395 \\\hline
\end{tabular}
\caption{Information about the datasets. The number of blocks depends on the hamming similarity threshold $h$($0.7$ used in this paper). The number of LD block-block (B-B) edges depends on the bianrization threshold $q$ (1.5\% for all the datasets except Parkinson, for which we used 0.25\% to speed up the BPM discovery in functional analysis II.)}
\label{tab:datasets} 
\end{table}

Three sources of existing knowledge are used for functional analysis: (a) a human functional network \cite{huttenhower2009exploring} \footnote{http://sonorus.princeton.edu/hefalmp/} (only edges with $w \geq 0.5$ are kept, where $w$ is the functional similarity between two genes ranging from $0$ to $1$), (b) human GO annotations \cite{ashburner2000gene} \footnote{Downloaded on Dec 29, 2010 from http://geneontology.org/} and (c) biological gene sets in the human molecular signature database (MSigDB\cite{subramanian2005gsa}) \footnote{Version 3.0 http://www.broadinstitute.org/gsea/msigdb/}

\subsection{Functional analysis I: Biological significance of the constructed genetic interaction network}

In this section, we investigate if the constructed human genetic interaction network has functional significance supported by independent biological data. Specifically, we check the functional similarity of the top LD block-block pairs in the genetic interaction network ($M_2$) and the similarity matrix $M_4$ with the human functional network ($M_5$) integrated in \cite{huttenhower2009exploring}. Because $M_5$ contains pair-wise gene relationships, a similar transformation as step 3 in Figure \ref{fig:flowchar_construct} is needed to derive an LD block-block functional similarity network ($M_6$). Similar to Equation \ref{eq:synergyBlock}, we summarize the gene-gene functional similarity network to an LD block-block functional similarity network with an aggregation function e.g. $max$, $mean$ etc. We used the $max$ function for the similar theoretical reasons and empirical observations as discussed for Equation \ref{eq:synergyBlock}. We note that there are many caveats with this choice but again, we keep the overall framework as the focus of this paper, and alternative aggregation functions will be explored further in future work.

\begin{figure}[t]
\centering
\subfigure[\small Parkinson bar plot\label{fig:xxx1}]{\includegraphics[width=.48\textwidth]{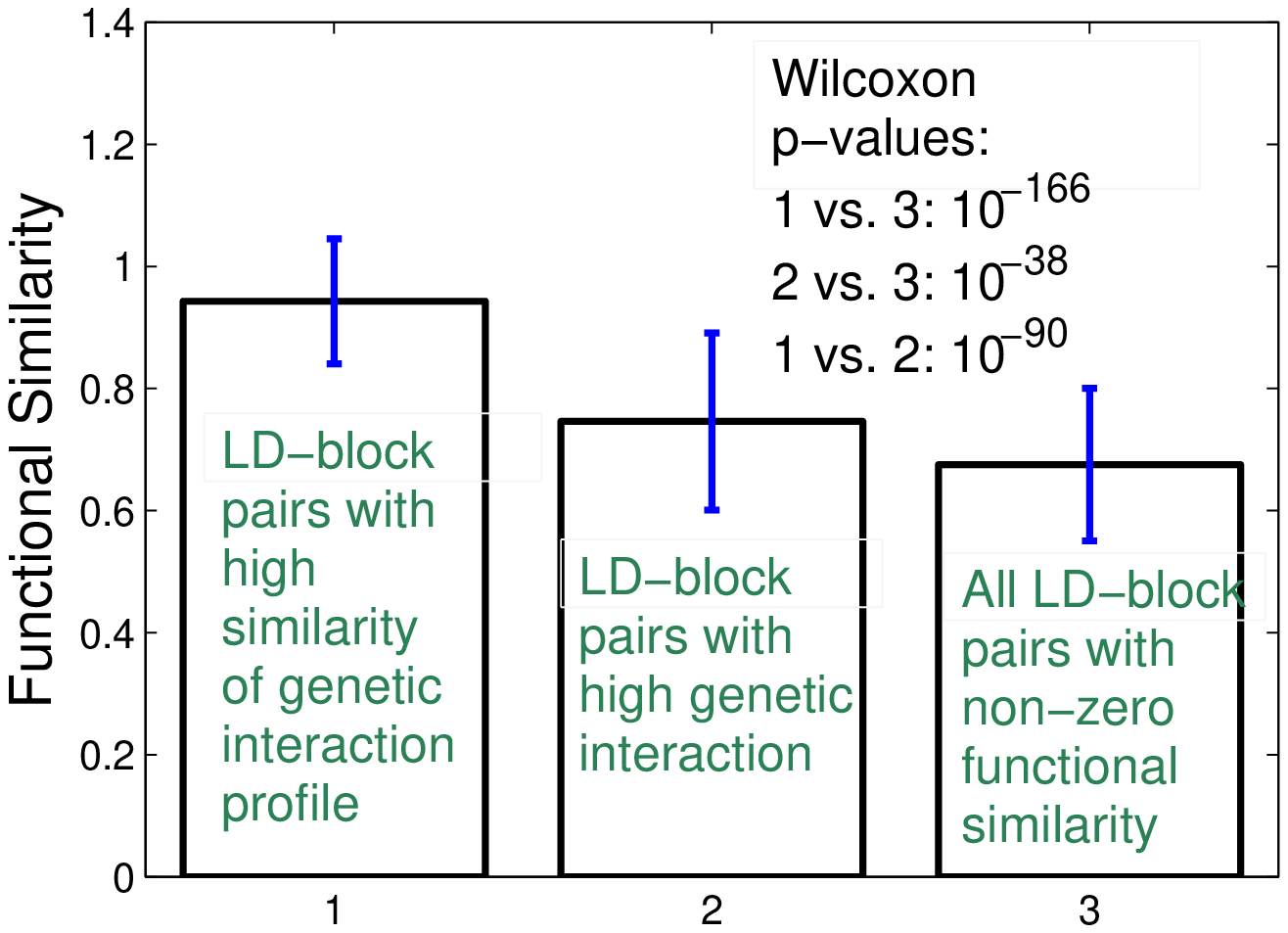}}
\subfigure[\small Parkinson scatter plot\label{fig:xxx2}]{\includegraphics[width=.48\textwidth]{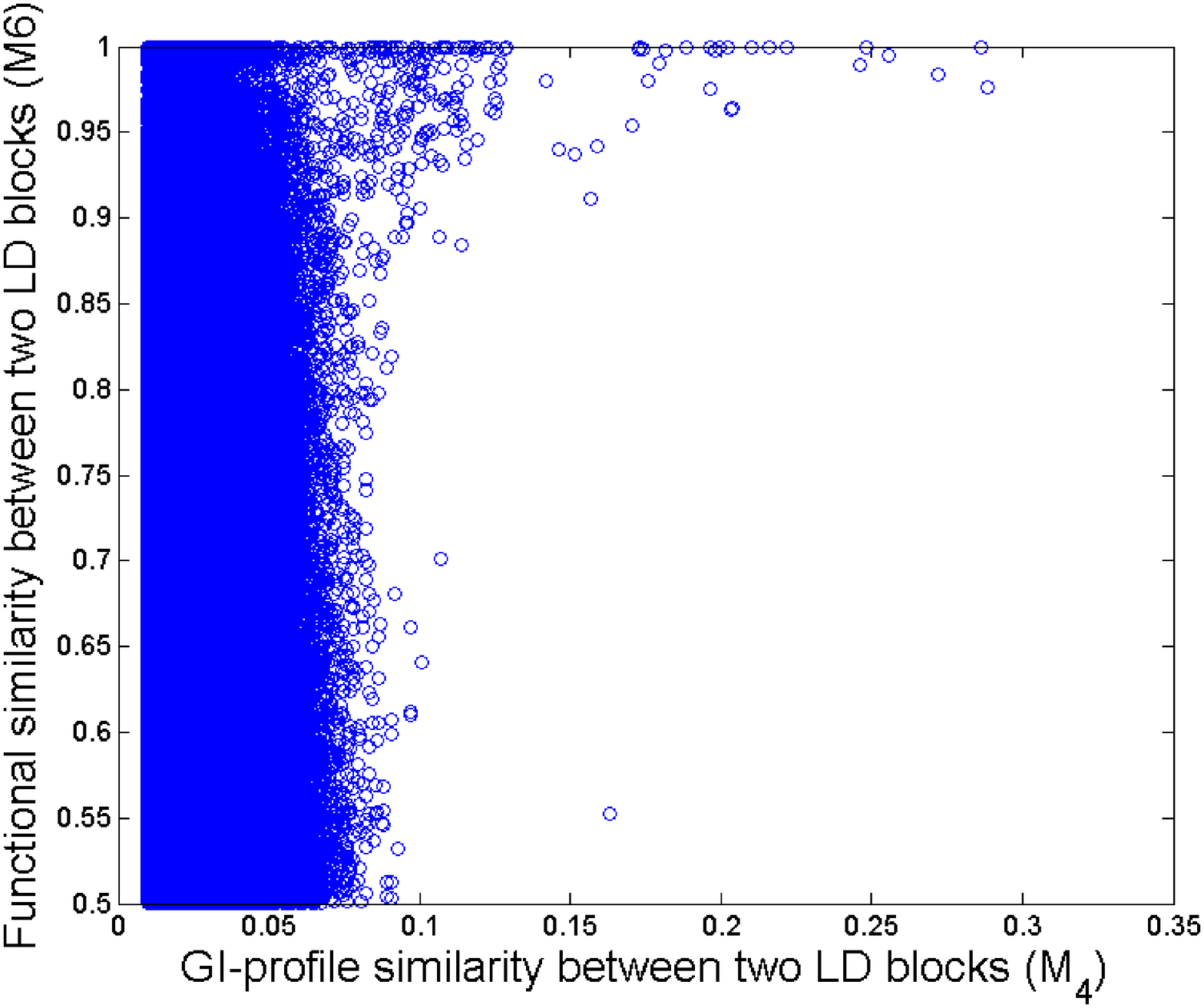}}
\subfigure[\small M-survival bar plot\label{fig:xxx3}]{\includegraphics[width=.48\textwidth]{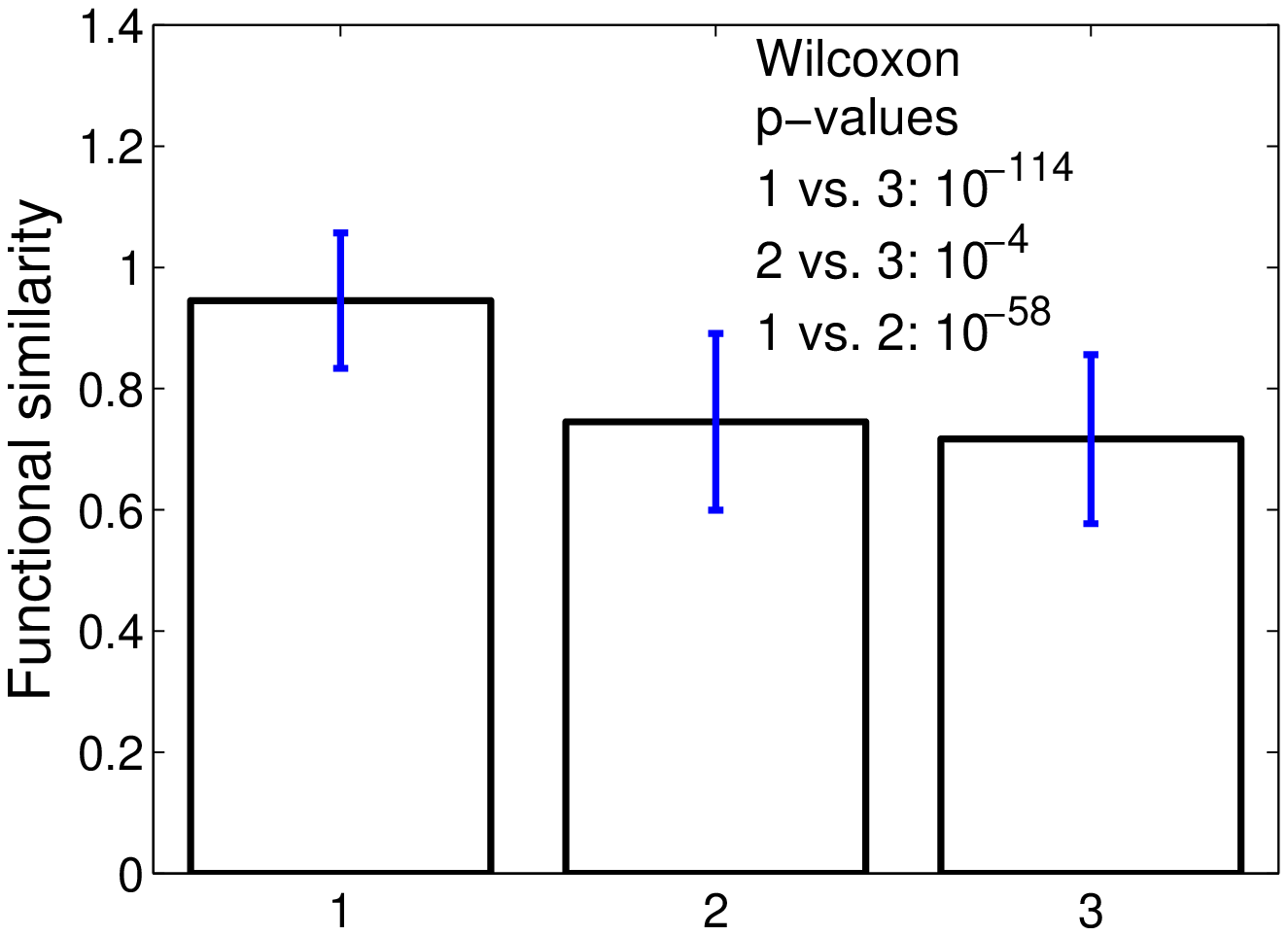}}
\subfigure[\small M-survival scatter plot\label{fig:xxx4}]{\includegraphics[width=.48\textwidth]{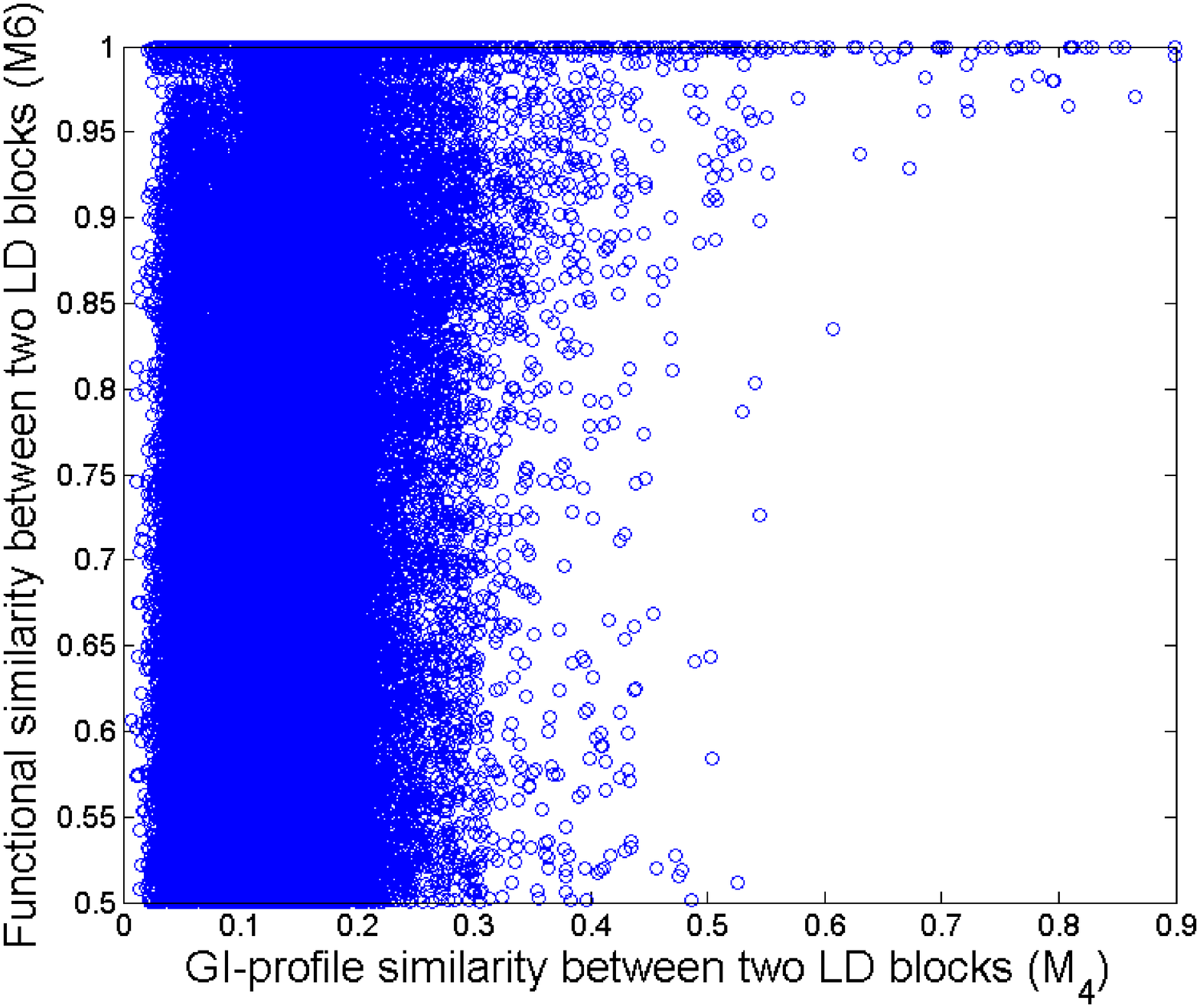}}
\caption{\small Functional similarity of top pairs of LD blocks in the genetic interaction network ($M_2$) and genetic interaction profile similarity network ($M_4$) on Parkinson and Myeloma-survival datasets. Each row shows two figures for each of the two datasets. Each subfigure on the left column shows three bars (with standard deviation also indicated on each), which tell the average functional similarity of three types of block pairs respectively: (1) those with high similarity of genetic interaction profile (top 300 edges in $M_4$), (2) block pairs with high genetic interaction (top 300 edges in $M_2$) and (3) all block pairs with functional similarity (baseline). The LD block-block pairs with zero functional similarity are not considered in any of the the three bars. We also include three p-values based on the Wilcoxon test on the functional similarities in the three groups. Each of the subfigures on the right column shows, for each pair of LD blocks, its genetic-interaction-profile similarity (in $M_4$) and its functional similarity (in $M_6$). The cluster of high functional similarities on the top-right corner corresponds to the significantly higher mean in the first bar in the subfigure on the left column of the same row.\vspace{-0.2in}}
\label{fig:fourAggregation}
\end{figure}

Figure \ref{fig:fourAggregation} shows the results on comparing the functional similarity of top pairs of LD blocks in genetic interaction network ($M_2$) and genetic interaction profile similarity network ($M_4$) with the human functional network. The following consistent observations can be made over all the six datasets \footnote{Two are shown in Fig. \ref{fig:fourAggregation}, and the others on the website}:

\begin{enumerate}
	\item \textbf{LD blocks with similar interaction profiles tend to contain functionally related genes} (Bar 1 is significantly higher than Bar 3): Specifically, if a pair of LD blocks have similar profile of genetic interactions, they tend to have significantly stronger functional relationship compared to the null distribution for all LD block-block pairs. This is biologically meaningful because similar genetic interaction profiles indicates probable common membership in a pathway or a protein complex, which agrees with observations on the yeast genetic interaction network \cite{costanzo2010genetic}.
	
	\item \textbf{Interacting LD blocks tend to contain functionally related genes} (Bar 2 is significantly higher than Bar 3): If a pair of LD blocks have high genetic interaction, they tend to have significantly strong functional relationship compared to the null distribution for all LD block-block pairs. This is biologically meaningful because high genetic interaction implies the functional complementarity between two pathways (or protein complexes). This also agrees with observations on the yeast genetic interaction network. 
	
	\item \textbf{Interaction profile similarity is more functionally specific than direct genetic interaction does} (Bar 1 is significantly higher than Bar 2): This indicates that, if a pair of LD blocks have similar profiles of genetic interactions, they tend to contain genes with stronger functional relationships compared to those that simply have a genetic interaction. This is expected because biological function is relatively more coherent for members of the same process or pathway than two complementary (genetically buffered) pathways. A similar trend has been observed in the yeast genetic interaction network. 
	
\end{enumerate}

These results suggest genetic interaction networks derived from human genetic association studies does produce networks with functional coherence, and this result is robust across the six studies we evaluated. We should note that the choice of $max$ over $mean$ for mapping the SNP-SNP network to LD block network and mapping the functional network to the LD block network does make a difference: the same result is not observed when choosing the $mean$, which suggests that the choice of how to summarize LD block-level statistics is a critical one. We note that other more sophisticated summary measures aside from the max will likely be superior but leave this for future \vspace{0.2in}work.

\subsection{Functional analysis II: Discovering and interpreting BPMs from the constructed genetic interaction network}
\label{sec:bpmresults}

In this subsection, we present the results of functional analysis II, i.e. BPM discovery and enrichment analysis.

Given a case-control dataset, a set of BPMs can be discovered from the corresponding binarized block-block genetic interaction network $M_3$. For each BPM, we compute the following measures to capture its various properties: 

\begin{enumerate}

	\item \textbf{Sizes}: We use three measures to describe the size of a BPM, $S_{sum}$(i.e. $|X|+|Y|$), $S_{max}$(i.e. $max(|X|,|Y|)$) and $S_{min}$(i.e. $min(|X|,|Y|)$), where $X$ and $Y$ are the two set of LD blocks in a BPM, as used in Equation \ref{eq:quasibclique}.
	
	\item \textbf{Genetic interaction density} as a quasi-biclique: We use three measures $D_{XX}$, $D_{XY}$, $D_{YY}$ which are the genetic interaction density of the submatrices $M_3(X,X)$, $M_3(X,Y)$ and $M_3(Y,Y)$, respectively. $D_{XY}$ says the genetic interaction density across the two sets of LD blocks, as an indicator of the strength of genetic interaction between the two sets. $D_{XX}$ and $D_{YY}$ tell the density within each of the two sets, which will be used to select a subset of BPMs with small density within the two sets as described in Section \ref{sec:bpm}.
	
	\item \textbf{Functional coherence} with respect to the human functional network: We use three measures $F_{XX}$, $F_{XY}$, $F_{YY}$ which are the average \footnote{Note that, here we use average to summarize the functional similarity between one set of LD blocks with another set of LD blocks (a BPM), where we expect a BPM to have strongly supporting functional relevance, thus we use average to enforce the relevance is not just from few block pairs. This is different from the $max$ function used to summarize the functional similarity between two LD blocks based on the genes that they are assigned with.} functional similarity in the submatrices $M_6(X,X)$, $M_6(X,Y)$ and $M_6(Y,Y)$, respectively. $F_{XX}$ and $F_{YY}$ indicate the two functional coherence within the two sets of LD blocks in a BPM, and $F_{XY}$ tells the functional relationship between the two set of LD blocks in a BPM.

\end{enumerate}

Because multiple BPMs can be discovered from a case-control dataset, corrections are needed for multiple hypothesis testing. Since the discovery is with respect to the genetic difference between cases and controls, we adopt the widely used permutation test \cite{subramanian2005gsa} in which the original case-control groups are randomly shuffled over the entire set of samples in a dataset (100 permutations were used on each of the datasets). For each of the permuted case-control grouping, the same pipeline for network construction and analysis as illustrated in Figures \ref{fig:flowchar_construct} and \ref{fig:flowchart_bpm_net_analysis} was repeated. Such permutations preserve the structure of the GWAS data and provide an unbiased, non-parametric correction for multiple hypothesis testing. After the 100 permutations, empirical false discovery rates (FDRs) (as used in \cite{subramanian2005gsa}) are computed for the following six BPM measures: $S_{sum}$, $S_{max}$, $D_{XY}$, $F_{XX}$, $F_{XY}$ and $F_{YY}$. Note that, $D_{XX}$, $D_{YY}$ are only used for filtering out uninteresting BPMs (threshold used: $0.35$), and we do not compute FDRs for them. As for $S_{sum}$, $S_{max}$ and $S_{min}$, we observe that they generally do not produce significant FDRs, and thus, it is not discussed in rest of this section but available on the supplementary website.

Two BPMs may overlap with each other on either of the two constituent LD block sets ($X$ and $Y$). Therefore, from the set of BPMs discovered from the real (or each permuted case-control grouping), we first select a subset of BPMs, by going through all the BPMs in decreasing order of $|X|$ and add a BPM to the subset if it has less than 50\% overlap on either the two sets ($X$ or $Y$) with all the BPMs that have been selected (initially empty). Note that, for the estimation of FDR for a BPM of size $|X|$ by $|Y|$ for one of the four measures ($D_{XY}$, $F_{XX}$, $F_{XY}$ and $F_{YY}$), the null distribution is only estimated from those random BPMs (discovered with permuted case-control groupings) that have sizes greater or equal to both $S_{max}$ and $S_{min}$. Such a size-specific estimation removes the bias of size difference that different BPMs may cause.

Table \ref{tab:fdrtable} shows the details of the number of BPMs discovered from each of the case-control datasets and the number of significant BPMs by each of the four measures (FDR $\leq 0.25$).

\begin{table}[!t]
\centering
\begin{tabular}{lllllll}\hline
&  $\sharp$ of  BPMs & & \multicolumn{4}{c}{ $\sharp$ of BPMs with}\\
&  discovered & Size of
&\multicolumn{4}{c}{each of the following FDRs $\leq 0.25$}\\
\cline{4-7}
Dataset & (non & Largest & FDR & FDR & FDR & FDR  \\
& redundant) & BPMs &  ($D_{XY}$) & ($F_{XX}$) & ($F_{YY}$) & ($F_{XY}$)  \\
\cline{1-7}
Parkinson & 20 & 19 by 4 & 7 & 3 & 4 & 7 \\
M-survival & 112 &18 by 4 & 11 & 4 & 4 & 3\\
Lung & 160 & 24 by 4 & 5 & 4 & 1 & 2  \\
Kidney & 65 & 15 by 4 & 1 & 1 & 2 & 2 \\
Myeloma & 38 & 17 by 4 & 6 & 5 & 4 & 7 \\
Bone & 171 & 13 by 5 & 3 & 5 & 1 & 1\\\hline
\end{tabular}
\caption{Number and sizes of the (significant) BPMs discovered from each dataset}
\label{tab:fdrtable} 
\end{table}

Several observations can be made from Table \ref{tab:fdrtable}.

\begin{enumerate}

	\item \textbf{Statistical significance of the discovered BPMs}: Many BPMs are significant with respect to the $D_{XY}$ on the the density of genetic interactions across the two sets. Specifically, there are significant BPMs discovered with respect to $D_{XY}$ on all the six datasets. This indicates that the existence of genetically buffered functional modules as captured by the BPM structure is evident in genome-wide case-control SNP datasets. These BPMs generally have larger sizes and interaction densities than the random BPMs discovered from permutation tests. 
	
	\item \textbf{Biological significance of the discovered BPMs}: Many BPMs are significant with respect to the last three measures, i.e. $F_{XX}$, $F_{YY}$ and $F_{XY}$, which suggests that they are not only statistically significant, but are also supported by independent genomic/proteomic evidence.  More specifically, the density of edges in the functional network, within both compensatory modules ($F_{XX}$, $F_{YY}$) as well as between them ($F_{XY}$), is frequently higher for the BPMs derived from the real data as compared to the random permutations of case-control groupings.

\end{enumerate}

\begin{figure}[t]
\centering
\includegraphics[width=.90\textwidth]{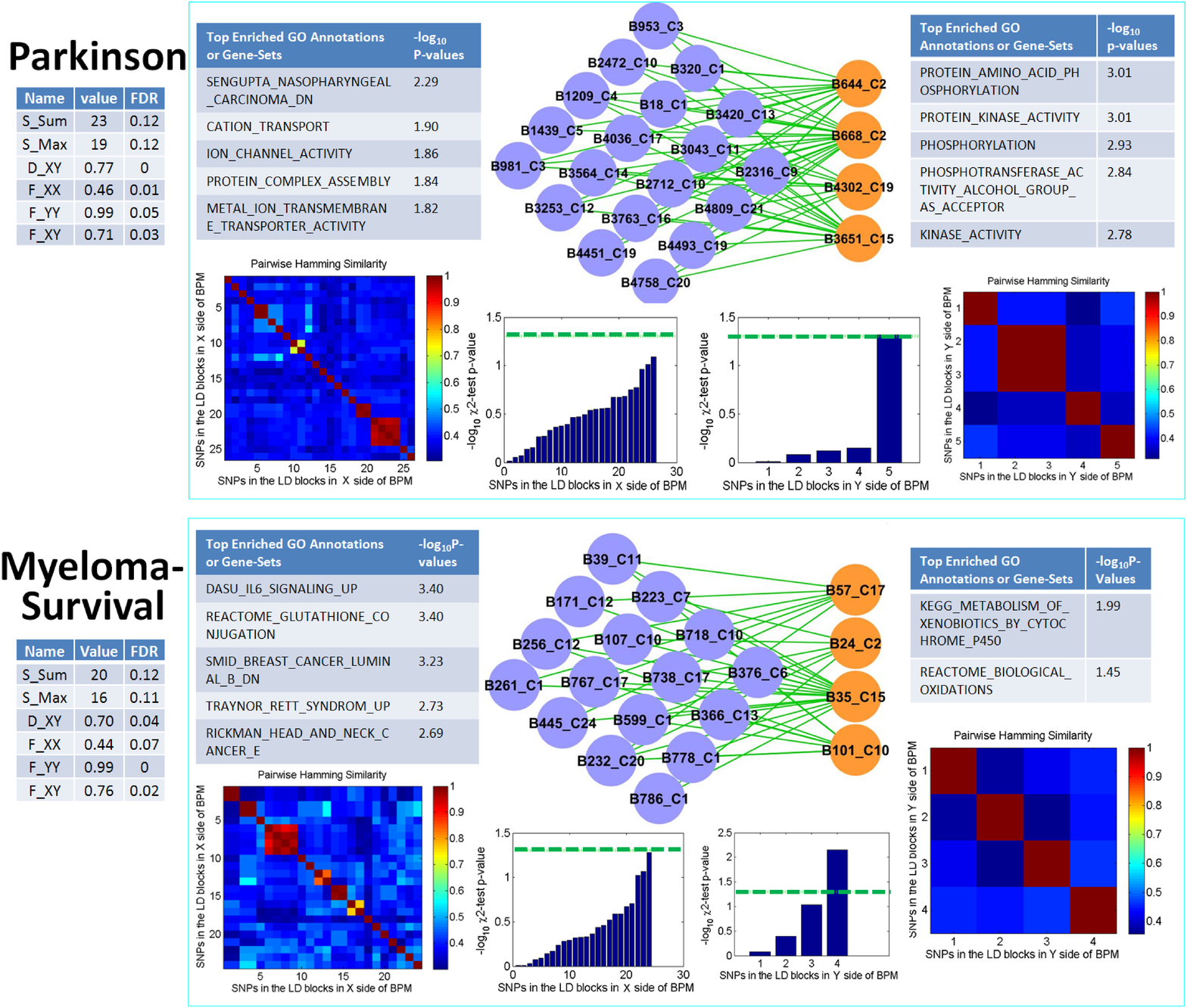}
\caption{Two illustrative BPMs discovered from Parkinson and Myeloma-survival. For each BPM, we display the LD block IDs on both sides and the chromosome that each block is located on. In addition, the seven measures and the corresponding FDRs are also shown for each BPM. Further, we also list the top five enriched GO annotations or MSigDB gene sets. A histogram of the $-log_{10}$ p-values ($\chi^2$ test) for all the individual SNPs in the blocks of each side of a BPM is also shown. A dashed line is shown in each histogram, which corresponds to the $-log_{10}(0.05)$ p-value without Bonferroni correction. All the SNPs have insignificant individual p-value supports the highlight of the proposed framework to discover significant BPMs containing SNPs with weak individual association, that are mostly ignored by existing approaches.} 
\label{fig:bpm_examples}
\end{figure}

While Table \ref{tab:fdrtable} gives an overall summary of the BPMs discovered from each of the datasets, Figure \ref{fig:bpm_examples} shows two illustrative examples of BPMs discovered from Parkinson and Myeloma-survival. Many BPMs from the other four datasets are available on the supplementary website.

Several observations can be made from Figure \ref{fig:bpm_examples}. We will use the BPM discovered from Parkinson as a running illustrative example of their interpretation.

\begin{enumerate}

	\item \textbf{The Statistical significance of the BPM discovered from Parkinson:} The table on the left shows the FDRs for each of the six BPM measures. As shown, the six FDRs are all below $0.15$ with several below $0.05$ (the latter four). A $D_{XY}$ of $0.73$ (FDR $\leq 0.01$) shows the dense genetic interactions between the two sets of LD blocks in this BPM. An $F_{YY}$ of 0.98 (FDR 0.05) indicates that the right set of LD blocks have strong LD block-block functional similarity, which agrees the concept of BPM, i.e. each side of the BPM is likely to be involved in a common pathway or a process. An $F_{XY}$ of $0.70$ (FDR 0.03) suggests that the two sets of LD blocks in this BPM may control two functions respectively, which may have compensatory function under genetic variations.
	
	\item \textbf{Most SNPs in the BPM have insignificant individual association}: The histograms of individual SNP association ($-log_{10}(\chi^2-test$ p-value)) indicate that, almost all of the SNPs in the significant BPM have insignificant individual p-value (below the green dashed line which corresponds to $-log_{10}(0.05)$ even before Bonferroni correction). This supports the utility of the proposed framework for discovering significant BPMs that contain SNPs with weak individual association, which would be mostly ignored by existing approaches.
	
	\item \textbf{All cross-set SNP pairs in the BPM have insignificant genetic interaction when considered separately}: We also compute FDR for each SNP pair in a BPM across the two sets of LD blocks. For the two examples shown in the figure, the lowest FDRs of individual SNP pairs are $0.45$ and $0.75$ in the Parkinson example and the Myeloma-survival example respectively. This indicates that none of the SNP pairs in the two datasets would be considered as significant epistasis if they are considered in an isolated manner. In contrast, discovering them as a BPM with the proposed approach yields their significant FDRs. This demonstrates the effectiveness of the proposed functional analysis for genome-wide case-control SNP datasets.
	
	\item \textbf{The two sets in the BPM contain LD blocks either from different chromosomes or from the same chromosome but with different genotype profiles}: The two heat maps of all pair's hamming similarities for the all the SNPs (grouped by LD blocks) indicate that the SNPs within each LD block have correlated genotype profiles while those from different LD blocks have different genotype profiles. This indicates that the BPM is not a trivial biclique that is due to LD structure. Note that, this is an illustration of the effectiveness and necessity of constructing and analyzing a block-level genetic interaction network instead of either a SNP-level network or a gene-level network as discussed in the methods section. The diversity of chromosomes in both sets of a BPM indicates that the two compensating functions (e.g. pathways) are from different chromosomes, showing the complexity of the mechanisms underlying the disease phenotype.

\end{enumerate}

\textbf{An illustrative interpretation on the BPM discovered on the Parkinson's dataset}: Given the statistical significance of the BPM structures discovered across many of the GWAS studies, we further asked whether the genes on both sides were enriched for known pathways or processes using gene sets defined by Gene Ontology terms or MSigDB. Consistent with their overlap with the functional network, a number of the modules involved in BPMs did show significant enrichment (see Figure \ref{fig:bpm_examples} for examples on the Parkinson and Myeloma-survival datasets or the supplementary website for a complete list). The BPM shown in Figure \ref{fig:bpm_examples} that is associated with Parkinson's disease was a pair of modules, one enriched for ion channel activity and the other enriched for protein kinase activity.  The ion channel activity enrichment is driven by three genes $ACCN1$, $KCNE1$, $TRPM3$, each of which comes from a separate LD block on three different chromosomes (chromosomes 9, 17, and 21, respectively).  This is potentially interesting as potassium channels were recently suggested as a possible new target for therapeutics \cite{wang2008potassium}.  It was hypothesized that such ion channels may affect the progressive loss of dopamine neurons, which is the main cause of Parkinson's disease.  It is also interesting to note that mouse knock-out mutants of $KCNE1$, a potassium voltage-gated channel protein, have been associated with the so-called shaker/waltzer phenotype, which is characterized by rapid bilateral circling during locomotion \cite{eugne2007developmental}.

The complementary module in this Parkinson's disease BPM was enriched for protein kinase activity due to the presence of the two protein kinases $MERTK$ (c-mer proto-oncogene tyrosine kinase) and $EIF2AK3$ (eukaryotic translation initiation factor 2-alpha kinase 3), suggesting that combined mutations affecting ion channel activity and one of these signaling pathways may be causal determinants of Parkinson's. While the specific link is unclear, it is interesting to note that $EIF2AK3$ is one of the key regulators of the eukaryotic translation initiation factor and thus, it controls global rates of protein synthesis in the cell \footnote{http://www.genecards.org/cgi-bin/carddisp.pl?gene=EIF2AK3}. It is certainly conceivable that mutations in such a protein with relatively global influence on protein levels could modify, and in this case aggravate, the effects of other mutations. In fact, mutations in another translation initiation factor, $EIF2B$, were recently associated with Vanishing White Matter disease, a disorder that causes rapid deterioration of the central nervous system \cite{leegwater2001subunits}.  Given the large number of genes involved in the LD blocks associated with each BPM, identifying the genes functionally responsible could be quite difficult and is one of the main caveats of this type of analysis. However, this process can be aided by simple enrichment analysis, which in this case appears to implicate processes whose link to Parkinson's disease seems plausible.

\section{Discussion}
\label{sec:discssion}

In this paper, we target the construction and functional analysis of disease-specific human genetic interaction networks from genome-wide association data designed for case-control studies on complex diseases. Specifically, we focused on genome-wide case-control SNP data, which has its linkage disequilibrium (LD) structure. We discussed the challenges in the detection of genetic interactions due to LD structure and propose a general approach with three steps: (1) estimating SNP-SNP genetic interactions, (2) identifying genome segments in linkage disequilibrium (LD) and mapping of SNP interactions to LD block-block interactions, and (3) mapping for LD blocks to genes. 

We performed two sets of functional analyses on six case-control SNP datasets to study if the constructed human genetic interaction network has functional significance supported by independent biological evidence by comparing with a human functional networks. We also demonstrated how the constructed interaction network can provide high-resolution insights about the compensation between pathways in their joint association with a disease phenotype by discovering between-pathway models. 

Comprehensive experimental results on six case-control datasets demonstrated that (i) From the perspective of genetic interaction analysis, the constructed human genetic interaction network has functional significance, and that biologically interesting motifs such as BPM that are common in lower eukaryotes also exist in the genetic interaction network discovered from human genetic variations associated with complex diseases such as cancers and Parkinson's disease; (ii) From the perspective of GWA data analysis, discovering BPMs from the constructed human genetic interaction network can help reveal novel biological insights about complex diseases, beyond existing approaches for GWAS data analysis that either ignore interactions between SNPs, or model different SNP-SNP genetic interactions separately rather than studying global genetic interaction networks as done in this study.

This paper focused on the presentation of the overall framework of constructing and analyzing human disease-specific genetic interaction network with GWAS data. There are a number of interesting and necessary directions for future work such as exploring the effect of different epistasis measures, the effect of different LD block identification approaches, the effect of different aggregation functions, and the effect of different gene mapping approaches. 

In conclusion, we want to highlight that, even though we chose to use some relatively simple and conservative options in the framework which needs further exploration as discussed above, the significant statistical and biological evidence obtained from the two sets of functional analyses demonstrate the effectiveness of the current framework in revealing several consistent observations over six case-control SNP datasets.

\end{document}